\documentstyle[twocolumn,aps,epsf]{revtex} 

\begin{document}
\draft                           


\title{Classical mechanism for negative magnetoresistance in two
dimensions}                           

\author{Alexander Dmitriev$^{1,2}$, Michel Dyakonov$^1$ and R\'emi
Jullien$^3$}  

\address{$^1$Laboratoire de Physique Math\'ematique$^\dagger$, 
Universit\'e 
         Montpellier 2,
         place E. Bataillon, 34095 Montpellier, France\\
         $^2$A. F. Ioffe Physico-Technical Institute, 194021 St.
Petersburg, 
         Russia\\ 
         $^3$Laboratoire des Verres$^\dagger$,  Universit\'e Montpellier
2,
         place E. Bataillon, 34095 Montpellier, France\\
$^\dagger$ Laboratoire associ\'e au Centre National de la Recherche
Scientifique (CNRS, France).}


\maketitle


\begin{abstract}
The classical two-dimensional problem of non-interacting electrons
scattered by a static impurity potential in the presence of magnetic
field is investigated both analytically and numerically. A strong
negative magnetoresistance exists in such a system, due to freely
circling electrons, which are not taken into account by the
Boltzmann-Drude approach. A parabolic magnetoresistance is found at low
fields.

\end{abstract}


------------------------------------------------------------------

The negative magnetoresistance, i.e. decrease of resistance in magnetic
field, frequently observed in semiconductors, as well as in metals,
remained a mystery for a long time, until Altshuler et al \cite{Aronov}
explained this phenomenon by quantum interference effects (weak
localization). Extensive experimental studies of magnetoresistance,
mostly in 2D semiconductor structures, have revealed that, apparently,
there are two distinct types of negative magnetoresistance: (i) a small
drop of resistivity observed at low fields, such that the classical
parameter $\beta=\omega_{c}\tau$ is small and (ii) a relatively large
(up to 50\%) decrease of resistivity at $\beta\ $\raise
-1.2mm\hbox{$\buildrel>\over\sim$}$\ 1$ or even at $\beta>>1$. Negative
magnetoresistance of type (ii) is observed before the onset of
Shubnikov-De Haas oscillations and may continue as a background trend
after the oscillations set in (see, for example, Refs. [2-5]. In some
instances, the small low-field dip of type (i) is superimposed on a
smooth overall decrease of resistivity\cite{wong}. Most of the studies
were devoted to the low-field magnetoresistance of type (i), which is
very well explained by the weak localization correction  for
non-interacting electrons \cite{Aronov}. As to the high-field effect
(ii), it is not so well understood, and is either attributed to the
effect of electron-electron interaction \cite{tsui}, which was
considered theoretically in Refs. \cite{Aronov1}, \cite{girvin}, or left
without any explanation. 

We recall that the simple Drude approach yealds zero magnetoresistance.
The Drude conductivity tensor is given by: 
$$\sigma_{xx} = {\sigma_{0}\over\ 1+\beta^2},\hspace{0.3in}\sigma_{xy} =
{\sigma_{0}\beta\over\ 1+\beta^2}, \eqno{(1)}$$
where $\sigma_{0}=ne^2\tau/m$ is the zero-field conductivity, $n$ is the
electron concentration, $e$ and $m$ are the electron charge and
effective mass respectively, $\tau$ is the momentum relaxation time,
$\beta =\omega_{c}\tau$, and $\omega_{c}=eB/mc$ is the electron
cyclotron frequency. For the resistivity tensor it follows:
$\rho_{xx}=\rho_{0}=1/\sigma_{0}$, $\rho_{xy}=\beta/\sigma_{0}=B/nec$,
and therefore the longitudinal resistivity is independent of the
magnetic field. This result applies to degenerate electrons for which
the time $\tau$, entering Eq. (1) should be taken at the Fermi energy
(for nondegenerate electrons one should take into account the dependence
of the scattering time $\tau$ on the electron energy, which, after
averaging Eqs. (1) over the Boltzmann energy distribution, results in a
{\it positive} magnetoresistance). 

In this report, we draw attention to a simple classical mechanism for
negative magnetoresistance. We consider non-interacting 2D electrons
with a given energy (the Fermi energy) scattered by short-range impurity
centers in the presence of a magnetic field perpendicular to the 2D
plane, and we show that for any type of scattering a strong negative
magnetoresistance should exist for  $\beta >>1$. We perform computer
simulations of the electron dynamics in such a system and find an
excellent agreement between the numerical results and a very simple
theory which is based on previously known results. Moreover we show for
the first time that the magnetoresistance is parabolic at low fields.
  
The problem was first studied in the pioneering work of Baskin et
al\cite{baskin}, and more recently by Bobylev et al \cite{boby}, who
considered specifically the 2D Lorentz model (scattering by hard disks)
and derived the main results relevant for the further discussion.

The main idea put forward in Refs.\cite{baskin}, \cite{boby} is that,
except for the case of small $\beta$, the Boltzmann-Drude approach does
not work, even as a first approximation, because of the existence of
"circling" electrons, which never collide with the short-range
scattering centers, the fraction of such electrons being\cite{boby}
$$P=\exp(-2 \pi R/\ell)=\exp(-2 \pi /\beta),\eqno{(2)}$$
where $R=v/\omega_{c}$ is the cyclotron radius, $v$ is the electron
(Fermi) velocity, and $\ell=v\tau$ is the electron mean free path.
Contrary to the assumption intrinsic to the Boltzmann-Drude approach, an
electron which happens to make one collisionless cycle will stay on its
cyclotron orbit forever. 
The behavior of the rest of electrons (the "wandering" electrons, in
terms of Ref. \cite{boby}), whose fraction is $1-P$, is controlled by
the parameter $NR^2$, the number of scatterers within the cyclotron
orbit, $N$ being the impurity concentration. For $NR^2>>1$ they behave
basically as predicted by the Drude theory, with an important
modification: after a collision with a given scatterer there is a
probability $P$ that the electron will recollide with the same scatterer
without experiencing any other collisions. As a result, for $\beta>>1$
the electron will recollide with the same impurity center many times,
and its trajectory will have a form of a rosette, sweeping a circular
area of radius $2R$ around the impurity center \cite{baskin}. Since the
number of impurities inside this area, $4\pi NR^2$, is large, eventually
the electron will collide with one of them, and thus continue its
diffusion in the 2D plane.  As it follows from the results of Ref.
\cite{boby}, frequent recollisons with the same center lead to the
isotropization of scattering, so that the effective $\tau$ in Eq. (1)
becomes field-dependent. This effect is absent if the scattering is
isotropic. 

At strong fields, when the parameter $NR^2$ becomes small enough, the
rosettes around different scatterers do not overlap anymore, the
colliding electrons become localized and give zero contribution to both
$\sigma_{xx}$ and $\sigma_{xy}$. This means that a percolation
transition should occur \cite{baskin}. The calculated threshold is
$(NR^2)_{c}=0.36$ \cite{boby}. 

Thus, there are two characteristic values of the magnetic field, $B_{1}$
defined by $\omega_{c}=1/\tau$ ($\beta=1$), and $B_{2}$ defined by 
$\omega_{c}=v\sqrt{N}$ ($NR^2=1$). The ratio
$B_{1}/B_{2}=(Nd^2)^{1/2}<<1$, where $d$ is the scattering
cross-section, is the small parameter of the theory.  

For the simpler case of isotropic scattering and $B<<B_{2}$ ($NR^2>>1$),
it follows from the results of Ref. \cite{boby} that the conductivity
tensor for wandering electrons is given simply by the conventional Drude
expressions, Eq. (1), with an additional factor $(1-P)$ in both
$\sigma_{xx}$ and $\sigma_{xy}$. The circling electrons behave like free
electrons with an effective concentration $nP$, giving a zero
contribution to  $\sigma_{xx}$, but contributing a term
$P\sigma_{0}/\beta =Pnec/B$ to $\sigma_{xy}$, and this is the reason why
the magnetoresistance is negative. This role of circling electrons was
overlooked in Ref. \cite{boby}, but was recognized later \cite{bobynew}
(see also Refs. \cite{kuz},\cite{basknew}).

Thus, the conductivity tensor is given by:
$$\sigma_{xx} = \sigma_{0}{1-P\over\ 1+\beta^2}, \eqno{(3a)}$$
$$\sigma_{xy} = \sigma_{0}\bigl ( (1-P){\beta\over\
1+\beta^2}+P{1\over\beta}\bigr ), \eqno{(3b)}$$
As a consequence, for the resistivity tensor we obtain
$$\rho_{xx}=\rho_{0}{1-P\over
1+P^2/\beta^2},\hspace{0.2in}\rho_{xy}=\rho_{0}\beta{1+P/\beta^2\over
1+P^2/\beta^2}.\eqno{(4)}$$
Formulas equivalent to Eqs. (3,4) were previously obtained by Baskin and
Entin \cite{basknew} for scattering by randomly positioned antidots. The
expression for $\rho_{xx}$ clearly exhibits negative magnetoresistance.
Since the terms $P/\beta^2$ and $P^2/\beta^2$ are small for any $\beta$,
Eqs. (4) are very similar to 
$$\rho_{xx}=\rho_{0}(1-P),\hspace{0.2in}\rho_{xy}=\rho_{0}\beta={B\over
nec},\eqno{(5)}$$
with an accuracy better than than 2\% for $\rho_{xx}$, and better than
4\% for $\rho_{xy}$. Note, that at low fields Eqs. (4,5) predict an
exponentially small magnetoresistance.

Before further discussion, let us present the results of our numerical
simulations. In our model a point particle (electron) with a given
absolute value of its velocity, v, is scattered by disks of diameter $d$
randomly positioned on a plane inside a square box of edge length $L$
(we take $L/d=1000$ to be sure that $L$ stays more than an order of
magnitude larger than the electronic mean-free path). Periodic boundary
conditions are imposed at the edges of the square box. Both the
hard-disk (Lorentz) model, which exhibits anisotropic scattering, and a
modified model with isotropic scattering are studied. To characterize
the coverage, we introduce a dimensionless concentration of scatterers
$c=\pi Nd^2/4$, which was changed from $c=0.025$ to $c=0.2$. Studies of
the percolation phenomena are beyond the scope of the present study. 

\begin{figure}

\hskip 15pt \epsfxsize=190pt{\epsffile{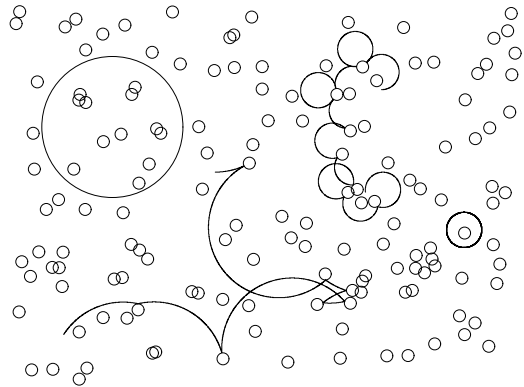}}
 
\vskip 15pt
 
\caption{Examples of simulated circling and wandering trajectories for
$\beta=1$ (left) and  $\beta=4$ (right) for a dimensionless impurity
concentration $c=\pi Nd^2/4=0.1$.  The actual fraction of circling
electrons at $\beta=1$ is very small.}
 
\end{figure}

In the simulation, we first choose an initial electron position at
random with an initial velocity along the $x$-direction. In a magnetic
field perpendicular to the plane the electron trajectory is made of
successive circular arcs of radius $R$. For each collision, we determine
the intersections of the trajectory with the disk periphery (the impact
point), which gives us the impact parameter, $b$, and calculate the
scattering angle, $\phi$, accordingly. We follow the electron velocities
$v_{x}(t)$ and $v_{y}(t)$ during a time $t=20\tau$ sufficient to get
reliable results for the integral below, and calculate the components of
the diffusion tensor by the standard formula:
 $$D_{ij} = {1\over 2}\int_0^\infty<v_{i}(0)v_{j}(t)>dt.\eqno{(6)}$$
For each value of field and disk concentration we take the average over
$10^2$ independent disk configurations, and, for each configuration,
over $10^6$ independent trials for the initial electron position. Of
course, at $B=0$, the trajectories are straight-line segments and
$D_{xy}$ should vanish (this provides a nice test for the numerical
precision). The conductivity tensor being proportional to the diffusion
tensor, the components of the resistivity $\rho_{ij}$ are calculated as
$ D_{ij}/(D_{xx}^2+D_{xy}^2)$, with an appropriate normalization.  For
the Lorentz model numerical calculations of this type were previously
performed \cite{kuz} with an emphasis on the percolation phenomenon.

Fig. 1 shows examples of simulated circling and wandering trajectories
for two values of magnetic field.
The circling electrons
give undamped oscillating contributions to the velocity correlation
functions in Eq. (6), accordingly, the integral in Eq. (6), strictly
speaking, does not converge at $t=\infty$, but is an oscillating
function of the upper limit. An average over these oscillations is
performed. The same result could be obtained if a small damping of these
oscillations were introduced (due, for example, to weak phonon
scattering).

The numerical results for $\rho_{xx}$ as a function of  $\beta$ for the
model with isotropic scattering are presented in Fig. 2 (top). The
resistivity is  normalized to the Boltzmann-Drude zero-field value,
$\rho_{0}$. The thick line is the theoretical curve predicted by Eq.
(4). One can see that the theoretical and numerical curves are
qualitatively similar and the quantitative agreement becomes better as
$c$ decreases. In the limit $c \rightarrow 0$ the numerical results
converge to the theoretical curve, as they should \cite{rem}.

\begin{figure}
 
\epsfxsize=215pt{\epsffile{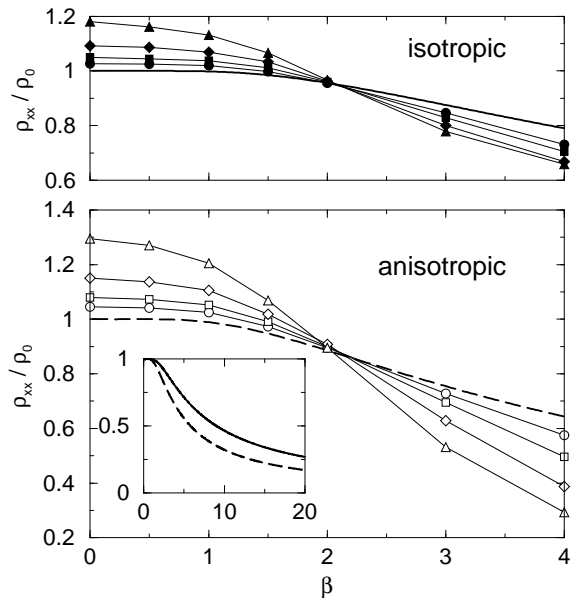}}
 
\caption{
Numerical results for the resistivity as a function of
$\beta=\omega_{c}\tau$ for different impurity concentrations, compared
to the theoretical curve given by Eq. (4) for the isotropic scattering
model (top) and  for the Lorentz model with anisotropic scattering
(bottom). Circles, squares, diamonds, and triangles correspond to
$c=0.025, 0.05, 0.1, 0.2$, respectively. The continuous and dashed thick
lines are the theoretical curves in the isotropic and anisotropic cases,
respectively, and they are depicted in the inset on a larger scale.
Note the surprising crossings at $\beta=2$.}
 
\end{figure}

Note that for finite $c$ the value of the zero-field resistivity is
higher than the Boltzmann value $\rho_{0}$. The relative correction for
small $c$ is proportional to $c\ln(1/c)$ and is due to recollisions with
the same impurity, which are not accounted for by the Boltzmann equation
\cite{hauge}. Note also, that the numerical results for finite $c$
approach the limiting theoretical curve from above for $\beta<2$, and
from below for $\beta>2$. This may be qualitatively explained as
follows. On the one hand, at small $\beta$ the resistivity for finite
$c$ is higher than the  $c \rightarrow 0$ Boltzmann value due to the
$c\ln(1/c)$ correction. We have found analytically that in magnetic
field this correction decreases quadratically in $\beta$, thus giving a
parabolic magnetoresistance proportional to $c\beta^2\sim 1/NR^2$. On
the other hand, at large $\beta$ we are on the way to the percolation
threshold, where $\rho_{xx}$ (but not  $\rho_{xy}$!) becomes zero. So,
obviously, for large $\beta$ and finite $c$ the resistivity should be
lower than the limiting value given by Eq. (4) \cite{rem}.

Fig. 2 (bottom) displays quite similar results obtained for the hard
disk Lorentz model (anisotropic scattering). The theoretical curve
(thick dashed line) was calculated using the results of Ref. \cite{boby}
for the wandering electrons and adding the contribution of circling
electrons, as explained above. In both cases all the numerical curves
for different $c$ cross the limiting theoretical curve at the same point
$\beta=2$ (within our numerical precision). We have no explanation for
this surprising finding so far.

One of the reasons, why the finite $c$ corrections are of interest, is
that Eq. (4) predicts exponentially small magnetoresistance for small
values of $\beta$. Corrections to this formula make the magnetic field
dependence parabolic and thus define the magnetoresistance at low
fields. In  order to isolate the $1/NR^2$ terms in magnetoresistance, we
look at the difference $$\delta= \bigl ({\rho_{xx}(B)-\rho_{xx}(0) \over
\rho_{xx}(0)}\bigr )_{num}-\bigl ({\rho_{xx}(B)-\rho_{xx}(0) \over
\rho_{xx}(0)}\bigr )_{th}. \eqno{(7)}$$
Here the second term in the right hand side is the normalized
magnetoresistance, given by Eq. (4) for isotropic scattering, or by a
similar formula taking care of the magnetic field dependence of $\tau$
for the anisotropic case. This term represents the limit
$NR^2\rightarrow 0$. The first term is the normalized magnetoresistance
found numerically.

This difference, as a function of $1/NR^2$ is presented in Fig. 3 for
both the isotropic and anisotropic scattering and for different values
of $c$. One can see that all the calculated points reasonably fit a
universal linear dependence, which corresponds to a quadratic dependence
on magnetic field. While such a dependence for small $\beta$ could be
anticipated, it is surprising that it persists for quite large values of
$\beta$. Thus at low magnetic field ($\beta<1$) the normalized
resistivity behaves like $1-0.15/NR^2$, the slope being deduced from
Fig. 3. For $\beta\ $\raise -1.2mm\hbox{$\buildrel>\over\sim$}$\ 1$ the
magnetoresistance is well described by Eqs. (4,5) or, for anisotropic
scattering, by similar formulas, which take care of the magnetic field
dependence of $\tau$.

\begin{figure}
 
\epsfxsize=230pt{\epsffile{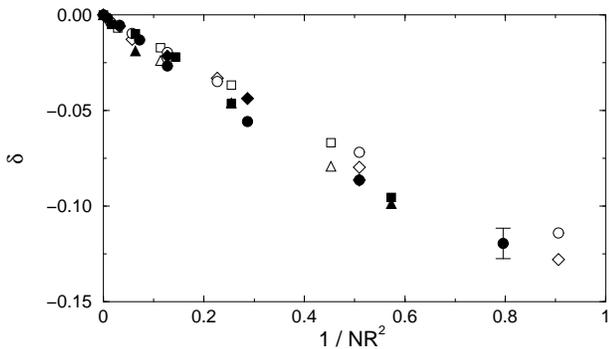}}
 
\caption{
The difference, $\delta$, between normalized magnetoresistances,
calculated in the limit $NR^2\rightarrow \infty$, and obtained numerically
(see Eq. (7)) for the isotropic (empty symbols) and anisotropic (filled
symbols) models, as a function of the square of the dimensionless field
$(B/B_{2})^2=1/NR^2$. Circles, squares, diamonds, and triangles
correspond to $c=0.025, 0.05, 0.1, 0.2$, respectively. The data fit a
linear dependence $\delta=-0.15/NR^2$.
}
\end{figure}

The classical approach is justified if the number of Landau levels below
the Fermi energy is large. However, it is irrelevant whether one can
describe the individual scattering events classically (when the
electron wave length is small compared to the size of the scatterer) or 
one needs a quantum-mechanical description (in the opposite case). So  
long as weak localisation corrections may be neglected, the differential 
scattering cross-section, even though calculated quantum-mechanically, may 
be used in the framework of a classical transport theory. We also remark 
that an infinite lifetime of the circling electrons is certainly an
idealization. In reality, even an electron whose orbit initialy avoids
the scattering centers, experiences forces which will gradually change
its trajectory. However, it is clear that, if the impurity potential
decreases fast enough compared to the average distance between
impurities, the basic features of the model will remain valid. The
situation is quite different for scattering by a long-range random
potential, which is typical for high-mobility 2D semiconductor
structures. Classical magnetotransport in this case was thoroughly
studied in Refs. \cite{shklovskii}, \cite{mirlin}. 

In summary, for short-range impurity scattering in two dimensions a
strong negative magnetoresistance exists, the main features of which
are  accounted for by a remarkably simple classical picture. There are
two species of electrons, wandering electrons which are described by the
conventional Drude theory, and circling electrons which are
collisionless and contribute only to the transverse conductivity. With
increasing magnetic field the fraction of circling electrons increases
at the expense of the wandering ones, leading to negative
magnetoresistance. However, as we have shown, at low fields the
magnetoresistance is entirely determined by small corrections to this
picture, which give a parabolic, rather than an exponentially small
negative magnetoresistance. This mechanism may provide a purely
classical explanation of, at least, some part of the experimental data
on negative magnetoresistance in two-dimensional structures.  

We thank W. Knap for useful discussions and for communicating his
experimental results prior to publication. We appreciate usefull
discussions with B. Shklovskii and D. Polyakov. We thank I. Gornyi for
bringing to our attention Refs. \cite{bobynew}, \cite{kuz},
\cite{basknew}.




\begin{references}
\bibitem{rem} According to the results of Ref. \cite{boby} for $c=0.2$
the percolation threshold should be reached at $\beta=3.3$ in the
isotropic case and at $\beta=2.5$ in the anisotropic case. However, in
our simulation we do not see any trace of the percolation transition for
these values of $\beta$. The reason may be that the threshold was
calculated assuming $d/R<<1$, while for $c=0.2$ and $\beta=3$ we have
$d/R\sim 1$. Presumably, lower values of $c$, and higher $\beta$ are
needed to obtain the theoretical threshold. Such a study is out of the
scope of the present work.
\end{references}
\end{document}